\documentclass{jors}

\definecolor{mygray}{gray}{0.6}
\renewcommand\headrule{}
\usepackage[backend=bibtex, firstinits=true, url=false, isbn=true]{biblatex}
\usepackage{amsmath}  % Basic
\usepackage{booktabs}  % Better tables
\usepackage{graphicx}  % Images
\usepackage{hyperref}  % Links etc
\usepackage{multicol}  % Multi column list
\usepackage{subcaption}  % Subfigures

\begin{document}
%% Set the header information
\pagestyle{fancy}
\definecolor{mygray}{gray}{0.6}
\renewcommand\headrule{}
\rhead{\footnotesize 3}
\rhead{\textcolor{gray}{UP JORS software Latex paper template version 0.1}}

\rule{\textwidth}{1pt}

\section*{(1) Overview}\label{sec:sec:overview}

\vspace{0.5cm}

\section*{Title}

An open framework for the reproducible study of the iterated prisoner's
dilemma.

\section*{Authors}

\begin{multicols}{2}
    \begin{enumerate}[noitemsep,topsep=0pt]
        % Core devs (alphabetic order of last name after Vince as corresponding
        % goes first).
        \item Knight, Vincent
        \item Campbell, Owen
        \item Harper, Marc
        \item Langner, Karol M.
        % All devs (alphabetic order of last name)
        \item Campbell, James
        \item Campbell, Thomas
        \item Carney, Alex
        \item Chorley, Martin
        \item Davidson-Pilon, Cameron
        \item Glass, Kristian
        \item Glynatsi, Nikoleta
        \item Ehrlich Tom{\'a}{\v s}
        \item Jones, Martin
        \item Koutsovoulos, Georgios
        \item Tibble, Holly
        \item M{\"u}ller, Jochen
        \item Palmer, Geraint
        \item Petunov, Piotr
        \item Slavin, Paul
        \item Standen, Timothy
        \item Visintini, Luis
        \item Molden, Karl
    \end{enumerate}
\end{multicols}

\section*{Paper Author Roles and Affiliations}

\begin{multicols}{2}
    \begin{enumerate}[noitemsep,topsep=0pt]
\item Development; Cardiff University
\item Development; Not affiliated
\item Development; Not affiliated
\item Development; Google Inc., Mountain View, CA
\item Development; Cardiff University
\item Development; St. Nicholas Catholic High School, Hartford
\item Development; Cardiff University
\item Development; Cardiff University
\item Development; Not affiliated
\item Development; Not affiliated
\item Development; Cardiff University
\item Development; Not affiliated
\item Development; Not affiliated
\item Development; The University of Edinburgh
\item Development; Not affiliated
\item Development; Not affiliated
\item Development; Cardiff University
\item Development; Not affiliated
\item Development; The University of Manchester
\item Development; Cardiff University
\item Development; Not affiliated
\item Development; Not affiliated
    \end{enumerate}
\end{multicols}

\section*{Abstract}

The Axelrod library is an open source Python package that allows for
reproducible game theoretic research into the Iterated Prisoner's Dilemma.
This area of research began in the 1980s but suffers from a lack of
documentation and test code. The goal of the library is to provide such
a resource, with facilities for the design of new strategies and
interactions between them, as well as conducting tournaments and ecological
simulations for populations of strategies.

With a growing collection of 136 strategies, the library is a also a
platform for an original tournament that, in itself, is of interest to the
game theoretic community.

This paper describes the Iterated Prisoner's Dilemma, the Axelrod library
and its development, and insights gained from some novel research.

\section*{Keywords}

Game Theory; Prisoners Dilemma; Python

\section*{Introduction}

Several Iterated Prisoner's Dilemma tournaments have generated much interest;
Axelrod's original tournaments \cite{Axelrod1980a,Axelrod1980b}, two 2004
anniversary tournaments \cite{kendall2007iterated}, and the Stewart and Plotkin
2012 tournament \cite{Stewart2012}, following the discovery of zero-determinant
strategies.  Subsequent research has spawned a number of papers (many of which
are referenced throughout this paper), but rarely are the results reproducible.
Amongst well-known tournaments, in only one case is the full original source code
available (Axelrod's second tournament \cite{Axelrod1980b}, in FORTRAN). In no
cases is the available code well-documented, easily modifiable, or released
with significant test suites.

To complicate matters further, a new strategy is often studied in isolation
with opponents chosen by the creator of that strategy. Often such strategies
are not sufficiently described to enable reliable recreation (in the absence of
source code), with \cite{slany2007some} being a notable counter-example. In
some cases, strategies are revised without updates to their names or published
implementations \cite{li2007design, li2011engineering}.
As such, the results cannot be reliably replicated and therefore have not met
the basic scientific criterion of falsifiability.

This paper introduces a software package: the Axelrod-Python library. The
Axelrod-Python project has the following stated goals:

\begin{itemize}[noitemsep,topsep=0pt]
    \item To enable the reproduction of Iterated Prisoner's Dilemma
    research as easily as possible
    \item To produce the de-facto tool for any future Iterated Prisoner's
    Dilemma research
    \item To provide as simple a means as possible for anyone to define and
    contribute new and original Iterated Prisoner's Dilemma strategies
\end{itemize}

The presented library is partly motivated by an ongoing discussion in the
academic community about reproducible research \cite{Crick2014a, Hong2015a,
Prlic2012, Sandve2013}, and is:

\begin{itemize}[noitemsep,topsep=0pt]
    \item Open: all code is released under an MIT license;
    \item Reproducible and well-tested: at the time of writing there is an excellent level of
        integrated tests with 99.73\% coverage (including property based tests:
        \cite{Hypothesis3.0.3})
    \item Well-documented: all features of the library are documented for ease of
        use and modification
    \item Extensive: 135 strategies are included, with infinitely-many
        available in the case of parametrised strategies
    \item Extensible: easy to modify to include new strategies and to run new
        tournaments
\end{itemize}

\section*{Review of the literature}\label{sec:review}

As stated in~\cite{Bendor1991}: ``\textit{few works in social science have had
the general impact of [Axelrod's study of the evolution of cooperation]}''.  In
1980, Axelrod wrote two papers:~\cite{Axelrod1980a,Axelrod1980b} which
describe a computer tournament that has been a major influence on
subsequent game theoretic work~\cite{Banks1990, Bendor1991, Boyd1987, Chellapilla1999,
DavidB1993, Doebeli2005, Ellison1994, Gotts2003, Hilbe2013, Isaac2008,
Kraines1989, Lee2015, Lorberbaum1994, Milgrom1982, Molander1985, Murnighan2015,
Press2012, Stephens2002, Stewart2012}. As described in~\cite{Bendor1991} this
work has not only had impact in mathematics but has also led to insights in
biology (for example in~\cite{Stephens2002}, a real tournament where Blue Jays
are the participants is described) and in particular in the study of evolution.

The tournament is based on an iterated game (see~\cite{Maschler2013} or similar
for details) where two players repeatedly play the normal form game
of~(\ref{equ:one-shot}) in full knowledge of each other's playing history to
date.  An excellent description of the \textit{one shot} game is given
in~\cite{Gotts2003} which is paraphrased below:

Two players must choose between \textit{Cooperate} (\(C\)) and \textit{Defect}
(\(D\)):

\begin{itemize}[noitemsep,topsep=0pt]
    \item If both choose \(C\), they receive a payoff of \(R\)
        (\textbf{R}eward);
    \item If both choose \(D\), they receive a payoff of \(P\)
        (\textbf{P}unishment);
    \item If one chooses \(C\) and the other \(D\), the defector receives a
        payoff of \(T\) (\textbf{T}emptation) and the cooperator a payoff of
        \(S\) (\textbf{S}ucker).
\end{itemize}

and the following reward matrix results from the Cartesian product of
two decision vectors $\langle C, D \rangle$,

\begin{equation}
    \begin{pmatrix}
        R,R & S,T\\
        T,S & P,P
    \end{pmatrix}\quad\text{such that } T>R>P>S \text{ and } 2R > T + S
    \label{equ:one-shot}
\end{equation}

The game of~(\ref{equ:one-shot}) is called the Prisoner's Dilemma. Specific
numerical values of \((R,S,T,P)=(3,0,5,1)\) are often used in the literature
\cite{Axelrod1980a, Axelrod1980b}, although any satisfying the conditions
in~\ref{equ:one-shot} will yield similar results. Axelrod's tournaments (and
further implementations of these) are sometimes referred to as Iterated
Prisoner's Dilemma (IPD) tournaments. An incomplete representative overview of
published tournaments is given in Table~\ref{tab:tournaments}.

\begin{table}[!hbtp]
    \begin{center}
        \begin{tabular}{ccccc}
            \toprule
            Year     & Reference                  & Number of Strategies & Type     & Source Code\\
            \midrule
            1979     & \cite{Axelrod1980a}        & 13                   & Standard & Not immediately available\\
            1979     & \cite{Axelrod1980b}        & 64                   & Standard & Available in FORTRAN\\
            1991     & \cite{Bendor1991}          & 13                   & Noisy    & Not immediately available\\
            2002     & \cite{Stephens2002}        & 16                   & Wildlife & Not a computer based tournament\\
            2005     & \cite{kendall2007iterated} & 223                  & Varied   & Not available \\
            2012     & \cite{Stewart2012}         & 13                   & Standard & Not fully available \\
            \bottomrule
        \end{tabular}
    \end{center}
    \caption{An overview of a selection of published tournaments. Not all
             tournaments were `standard' round robins; for more details
             see the indicated references.}\label{tab:tournaments}
\end{table}

In \cite{Milgrom1982} a description is given of how incomplete information can
be used to enhance cooperation, in a similar approach to the proof of the Folk
theorem for repeated games \cite{Maschler2013}. This aspect of incomplete
information is also considered in \cite{Bendor1991, Lee2015, Molander1985} where
``noisy'' tournaments randomly flip the choice made by a given strategy. In
\cite{Murnighan2015}, incomplete information is considered in the sense of a
probabilistic termination of each round of the tournament.

As mentioned before, IPD tournaments have been studied in an evolutionary
context: \cite{Ellison1994, Lee2015, Press2012, Stewart2012} consider this in a
traditional evolutionary game theory context. These works investigate
particular evolutionary contexts within which cooperation can evolve and
persist. This can be in the context of direct interactions between strategies
or population dynamics for populations of many players using a variety of
strategies, which can lead to very different results. For example, in
\cite{Lee2015} a machine learning algorithm in a population context outperforms
strategies described in \cite{Press2012} and \cite{Stewart2012} that are
claimed to dominate any evolutionary opponent in head-to-head interactions.

Further to these evolutionary ideas, \cite{Chellapilla1999, DavidB1993} are
examples of using machine learning techniques to evolve particular strategies.
In \cite{Axelrod}, Axelrod describes how similar techniques are used to
genetically evolve a high performing strategy from a given set of strategies.
Note that in his original work, Axelrod only used a base strategy set of 12
strategies for this evolutionary study. This is noteworthy as
the library now boasts over 136 strategies that are
readily available for a similar analysis.

\section*{Implementation and architecture}

\section*{Description of the Axelrod Python package}\label{sec:description-of-axelrod-python}

The library is written in Python (\url{http://www.python.org/}) which is a
popular language in the academic community with libraries developed for a
variety of uses including:

\begin{itemize}[noitemsep,topsep=0pt]
    \item Algorithmic Game Theory \cite{Mckelvey06gambit:software}
          (\url{http://gambit.sourceforge.net/})).
    \item Astrophysics \cite{astropy} (\url{http://www.astropy.org/});
    \item Data manipulation \cite{mckinney-proc-scipy-2010}
          (\url{http://pandas.pydata.org/});
    \item Machine learning \cite{scikit-learn} (\url{http://scikit-learn.org/});
    \item Mathematics \cite{sage} (\url{http://www.sagemath.org/});
    \item Visualisation \cite{Hunter:2007} (\url{http://matplotlib.org/});
\end{itemize}

Furthermore, in \cite{Isaac2008} Python is described as an appropriate language for the
reproduction of Iterated Prisoner's Dilemma tournaments due to its object
oriented nature and readability.

The library itself is available at
\url{https://github.com/Axelrod-Python/Axelrod}. This is a hosted git
repository. Git is a version control system which is one of the
recommended aspects of reproducible research \cite{Crick2014a, Sandve2013}.

As stated in the \textbf{Introduction}, one of the main goals of the library is
to allow for the easy contribution of strategies. Doing this requires the
writing of a simple Python class (which can inherit from other predefined
classes). All components of the library are automatically tested using a
combination of unit, property and integration tests. These tests are run as new
features are added to the library to ensure compatibility (they are also run
automatically using \url{travis-ci.org}). When submitting a strategy, a
simple test is required which ensures the strategy behaves as expected.  Full
contribution guidelines can be found in the documentation, which is also part of
the library itself and is hosted using \url{readthedocs.org}.  As an example,
Figures~\ref{fig:grudger} and~\ref{fig:grudger_test} show the source code for
the Grudger strategy as well as its corresponding test.

\begin{figure}[!hbtp]
    \begin{verbatim}
class Grudger(Player):
    """A player starts by cooperating however will defect if
       at any point the opponent has defected."""

    name = 'Grudger'
    classifier = {
        'memory_depth': float('inf'),  # Long memory
        'stochastic': False,
        'inspects_source': False,
        'manipulates_source': False,
        'manipulates_state': False
    }

    def strategy(self, opponent):
        """Begins by playing C, then plays D for the remaining
           rounds if the opponent ever plays D."""
        if opponent.defections:
            return D
        return C
    \end{verbatim}
    \caption{Source code for the Grudger strategy.}
    \label{fig:grudger}
\end{figure}

\begin{figure}[!hbtp]
    \begin{verbatim}
class TestGrudger(TestPlayer):

    name = "Grudger"
    player = axelrod.Grudger
    expected_classifier = {
        'memory_depth': float('inf'),  # Long memory
        'stochastic': False,
        'inspects_source': False,
        'manipulates_source': False,
        'manipulates_state': False
    }

    def test_initial_strategy(self):
        """
        Starts by cooperating
        """
        self.first_play_test(C)

    def test_strategy(self):
        """
        If opponent defects at any point then the player will defect forever
        """
        self.responses_test([C, D, D, D], [C, C, C, C], [C])
        self.responses_test([C, C, D, D, D], [C, D, C, C, C], [D])
    \end{verbatim}
    \caption{Test code for the Grudger strategy.}
    \label{fig:grudger_test}
\end{figure}

\begin{figure}[!hbtp]
	\centering
	\includegraphics[width=.9\textwidth]{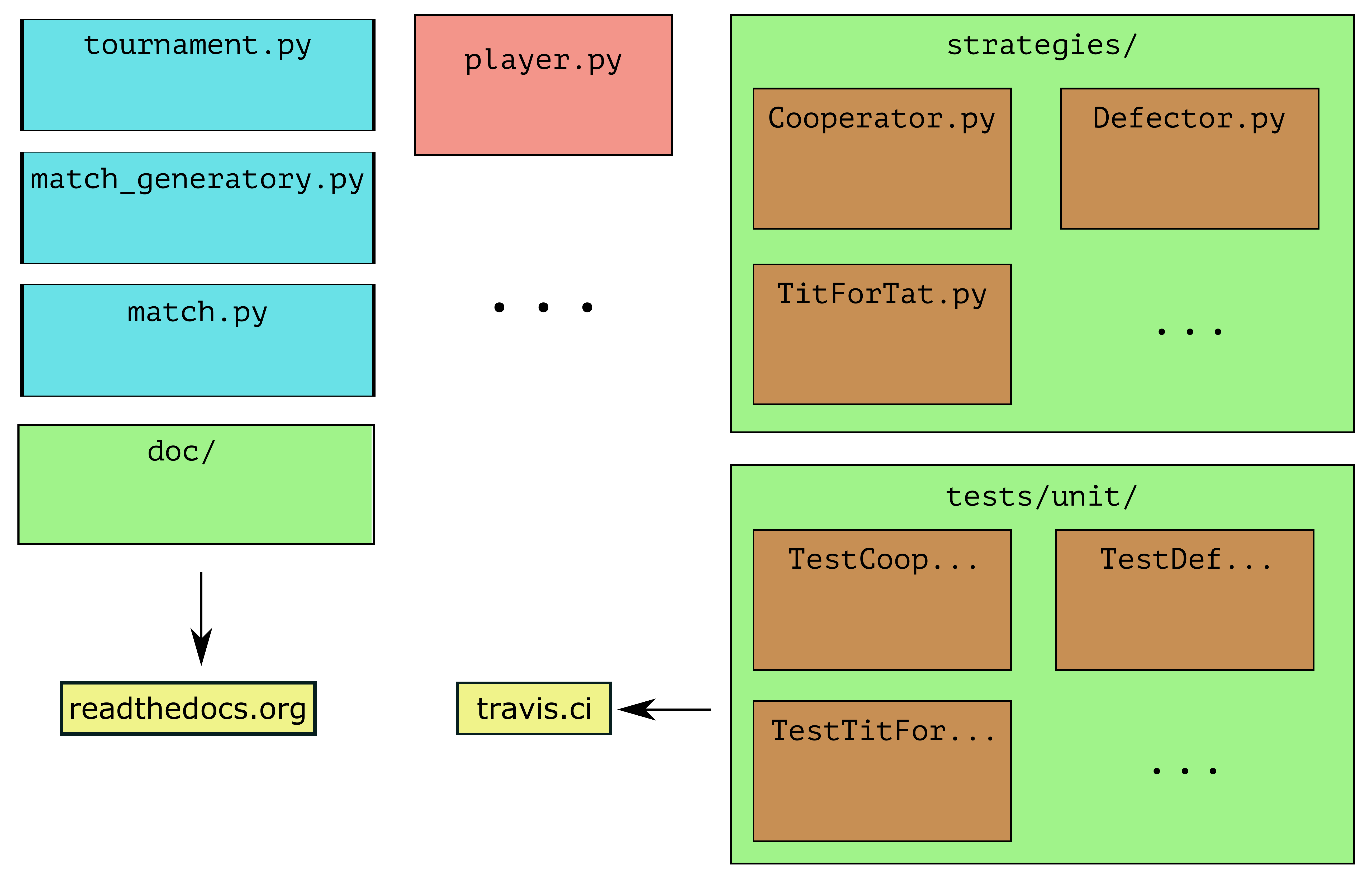}
	\caption{An overview of the source code.}
	\label{fig:overview}
\end{figure}

You can see an overview of the structure of the source code in
Figure~\ref{fig:overview}. This shows the parallel collection of strategies and
their tests. Furthermore the underlying engine for the library is a class for
tournaments which lives in the \texttt{tournament.py} module. This class is
responsible for coordinating the play of generated matches (from the
\texttt{match.py} module). This generation of matches is the responsibility of a
\textit{match generator} class (in the \texttt{match\_generator.py} module)
which is designed in such a way as to be easily modifiable to create new types
of tournaments. This is described further in a tutorial in the documentation
which shows how to easily create a tournament where players only play each other
with probability 0.5.  This will be discussed further in the reuse section of
this paper.

To date the library has had contributions from 26 contributors from a variety of
backgrounds which are not solely academic. These contributions have been mostly
in terms of strategies. One strategy is the creation of an undergraduate
mathematics student with little prior knowledge of programming. Multiple other
strategies were written by a 15 year old secondary school student. Both of these
students are authors of this paper. As well as these strategy contributions,
vital architectural improvements to the library itself have also been
contributed.

\section*{(2) Availability}\label{sec:availability}

\section*{Operating system}

The Axelrod library runs on all major operating systems: Linux, Mac OS X and
Windows.

\section*{Programming language}

The library is continuously tested for compatibility with Python 2.7 and the two
most recent python 3 releases.

\section*{Additional system requirements}

There are no specific additional system requirements.

\section*{Support}

Support is readily available in multiple forms:

\begin{itemize}[noitemsep,topsep=0pt]
    \item An online chat channel:
        \url{https://gitter.im/Axelrod-Python/Axelrod}.
    \item An  email group:
        \url{https://groups.google.com/forum/#!topic/axelrod-python}.
\end{itemize}

\section*{Dependencies}

The following Python libraries are required dependencies:

\begin{multicols}{2}
    \begin{itemize}[noitemsep,topsep=0pt]
        \item Numpy 1.9.2
        \item Matplotlib 1.4.2 (only a requirement if graphical output is
            required)
        \item Tqdm 3.4.0
        \item Hypothesis 3.0 (only a requirement for development)
    \end{itemize}
\end{multicols}

\section*{List of contributors}

The names of all the contributors are not known: as these were mainly done
through Github and some have not provided their name or responded to a
request for further details. Here is an incomplete list:

\begin{multicols}{2}
    \begin{itemize}[noitemsep,topsep=0pt]
        \item {Owen Campbell}
        \item {Marc Harper}
        \item {Vincent Knight}
        \item {Karol M. Langner}
        \item {James Campbell}
        \item {Thomas Campbell}
        \item {Alex Carney}
        \item {Martin Chorley}
        \item {Cameron Davidson-Pilon}
        \item {Kristian Glass}
        \item {Nikoleta Glynatsi}
        \item {Tom{\'a}{\v s} Ehrlich}
        \item {Martin Jones}
        \item {Georgios Koutsovoulos}
        \item {Holly Tibble}
        \item {Jochen M{\"u}ller}
        \item {Geraint Palmer}
        \item {Paul Slavin}
        \item {Timothy Standen}
        \item {Luis Visintini}
        \item {Karl Molden}
        \item {Jason Young}
        \item {Andy Boot}
        \item {Anna Barriscale}
    \end{itemize}
\end{multicols}

\section*{Software location:}

{\bf Archive}

\begin{description}[noitemsep,topsep=0pt]
    \item[Name:] Zenodo
    \item[Persistent identifier:]
\href{https://zenodo.org/record/55509}{10.5281/zenodo.55509}
    \item[Licence:] MIT
    \item[Publisher:]  Vincent Knight
    \item[Version published:] Axelrod: 1.2.0
    \item[Date published:] 2016-06-13
\end{description}

{\bf Code repository}

\begin{description}[noitemsep,topsep=0pt]
    \item[Name:] Github
    \item[Identifier:] \url{https://github.com/Axelrod-Python/Axelrod}
    \item[Licence:] MIT
    \item[Date published:] 2015-02-16
\end{description}

\section*{Reuse potential}\label{sec:reuse}

The Axelrod library has been designed with sustainable software practices in
mind. There is an extensive documentation suite:
\url{axelrod.readthedocs.org/en/latest/}. Furthermore, there is a growing set
of example Jupyter notebooks available here:
\url{https://github.com/Axelrod-Python/Axelrod-notebooks}.

The availability of a large number of strategies makes this
tool an excellent and obvious example of the benefits of open research which
should positively impact the game theory community.
This is evidently true already as the library has been used to study and create
interesting and powerful new strategies.

Installation of the library is straightforward via standard python installation
repositories (\url{https://pypi.python.org/pypi}). The package name is
\texttt{axelrod} and can thus be installed by calling: \texttt{pip install
axelrod} on all major operating systems (Windows, OS X and Linux).

\begin{figure}[!hbtp]
	\centering
	\includegraphics[width=.75\textwidth]{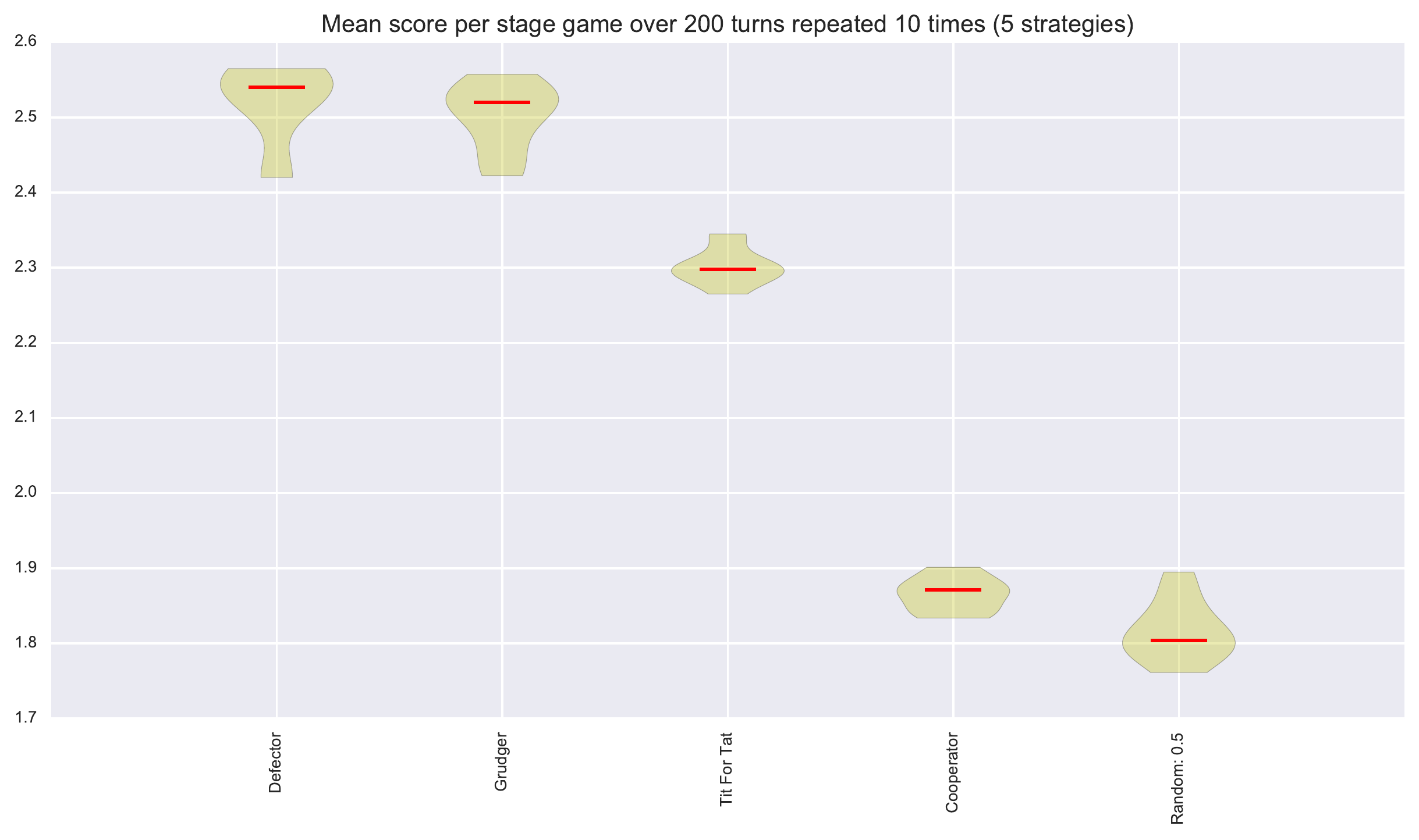}
	\caption{The results from a simple tournament.}
	\label{fig:demo_tournament}
\end{figure}

Figure~\ref{fig:demo_tournament_commands} shows a very simple example of using
the library to create a basic tournament giving the graphical output shown in
Figure~\ref{fig:demo_tournament}.

\begin{figure}[!hbtp]
    \begin{verbatim}
>>> import axelrod
>>> strategies = [s() for s in axelrod.demo_strategies]
>>> tournament = axelrod.Tournament(strategies)
>>> results = tournament.play()
>>> plot = axelrod.Plot(results)
>>> p = plot.boxplot()
>>> p.show()
    \end{verbatim}
    \caption{A simple set of commands to create a demonstration tournament. The
        output is shown in Figure~\ref{fig:demo_tournament}.}
    \label{fig:demo_tournament_commands}
\end{figure}

\section*{New strategies, tournaments and implications}\label{sec:new-strategies-and-implications}

\begin{figure}[!hbtp]
	\centering
	\includegraphics[width=.75\textwidth]{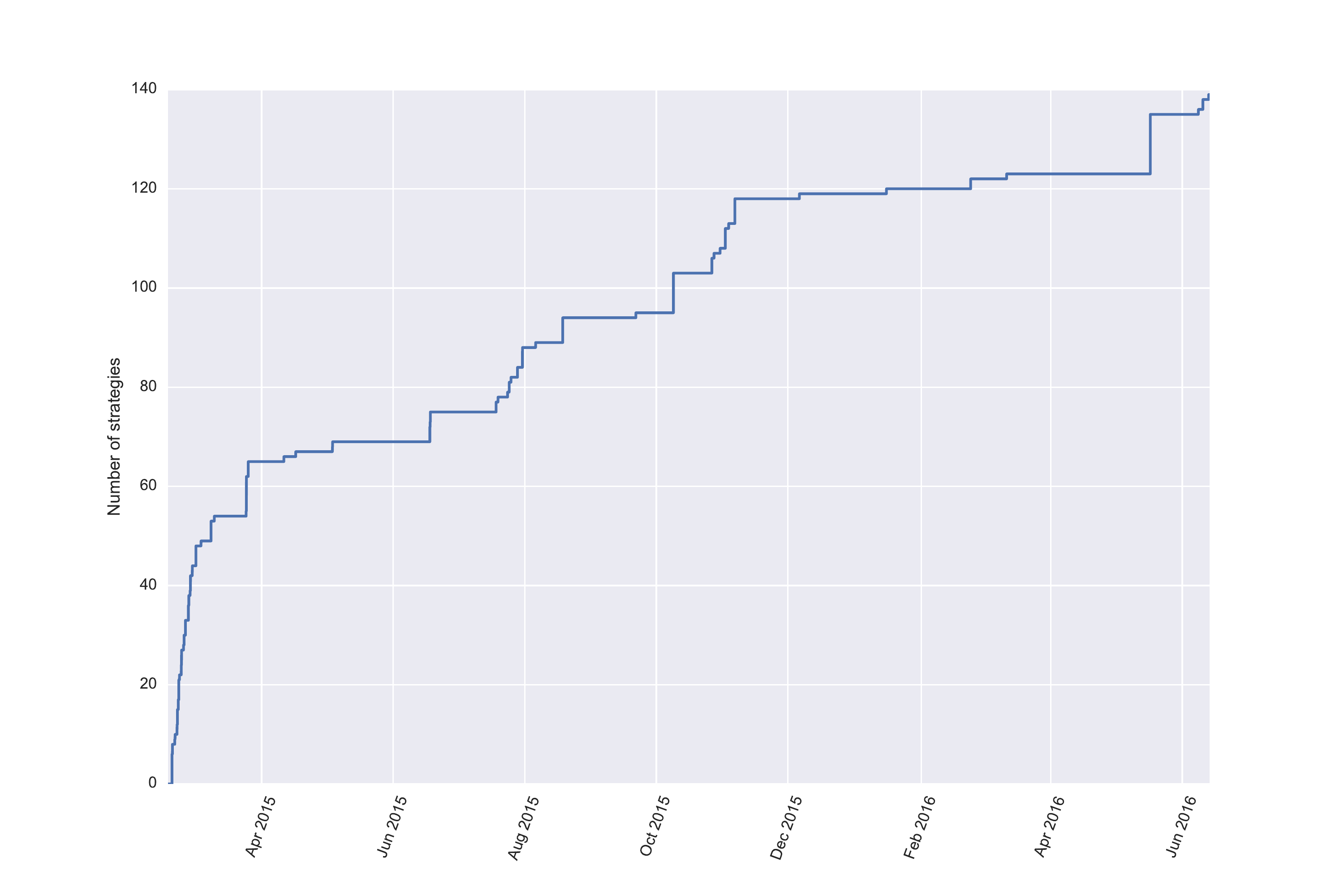}
	\caption{The number of strategies included in the library}
	\label{fig:number_of_strategies_against_date}
\end{figure}

Due to the open nature of the library the number of strategies included has
grown at a fast pace, as can be seen in
Figure~\ref{fig:number_of_strategies_against_date}.

Nevertheless, due to previous research being done in an irreproducible manner
with, for example, no source code and/or vaguely described strategies, not all
previous tournaments can yet be reproduced. In fact, some of the early
tournaments might be impossible to reproduce as the source code is apparently
forever lost. This library aims ensure reproducibility in the future.

One tournament that is possible to reproduce is that of
\cite{Stewart2012}. The strategies used in that tournament are the following:

\begin{multicols}{2}
    \begin{enumerate}[noitemsep,topsep=0pt]
        \item Cooperator
        \item Defector
        \item ZD-Extort-2
        \item Joss: 0.9
        \item Hard Tit For Tat
        \item Hard Tit For 2 Tats
        \item Tit For Tat
        \item Grudger
        \item Tit For 2 Tats
        \item Win-Stay Lose-Shift
        \item Random: 0.5
        \item ZD-GTFT-2
        \item GTFT: 0.33
        \item Hard Prober
        \item Prober
        \item Prober 2
        \item Prober 3
        \item Calculator
        \item Hard Go By Majority
    \end{enumerate}
\end{multicols}

\begin{figure}[!hbtp]
	\centering
	\includegraphics[width=.75\textwidth]{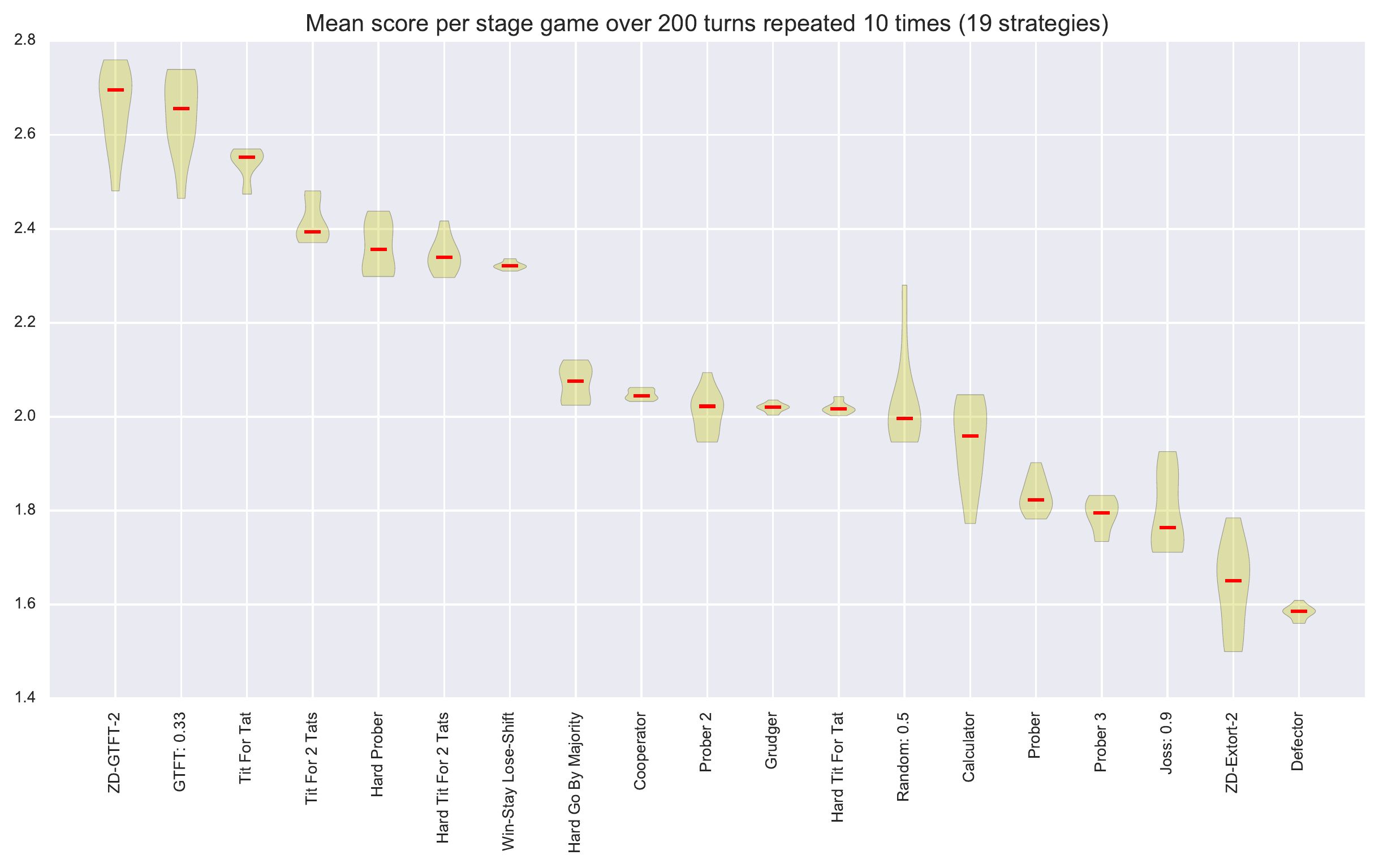}
	\caption{The results from \cite{Stewart2012}.}
	\label{fig:stewart_tournament}
\end{figure}

This can be reproduced as shown in Figure~\ref{fig:stewart-code}, which gives
the plot of Figure~\ref{fig:stewart_tournament}. Note that slight differences
with the results of \cite{Stewart2012} are due to stochastic behaviour of some
strategies.

\begin{figure}[!hbtp]
    \begin{verbatim}
>>> import axelrod
>>> strategies = [axelrod.Cooperator(),
...               axelrod.Defector(),
...               axelrod.ZDExtort2(),
...               axelrod.Joss(),
...               axelrod.HardTitForTat(),
...               axelrod.HardTitFor2Tats(),
...               axelrod.TitForTat(),
...               axelrod.Grudger(),
...               axelrod.TitFor2Tats(),
...               axelrod.WinStayLoseShift(),
...               axelrod.Random(),
...               axelrod.ZDGTFT2(),
...               axelrod.GTFT(),
...               axelrod.HardProber(),
...               axelrod.Prober(),
...               axelrod.Prober2(),
...               axelrod.Prober3(),
...               axelrod.Calculator(),
...               axelrod.HardGoByMajority()]
>>> tournament = axelrod.Tournament(strategies)
>>> results = tournament.play()
>>> plot = axelrod.Plot(results)
>>> p = plot.boxplot()
>>> p.show()
    \end{verbatim}
    \caption{Source code for reproducing the tournament of \cite{Stewart2012}}
    \label{fig:stewart-code}
\end{figure}

In parallel to the Python library, a tournament is being kept up to date that
pits all available strategies against each other. Figure~\ref{fig:tournament}
shows the results from the full tournament which can also be seen (in full
detail) here: \url{http://axelrod-tournament.readthedocs.org/}. Data sets are
also available showing the plays of every match that takes place. Note that to
recreate this tournament simply requires changing a single line of the code
shown in Figure~\ref{fig:demo_tournament_commands}, changing:

\begin{verbatim}
>>> strategies = [s() for s in axelrod.demo_strategies]}
\end{verbatim}

to:

\begin{verbatim}
>>> strategies = [s() for s in axelrod.ordinary_strategies]}.
\end{verbatim}

\begin{figure}[!hbtp]
    \centering
    \includegraphics[width=.75\textwidth]{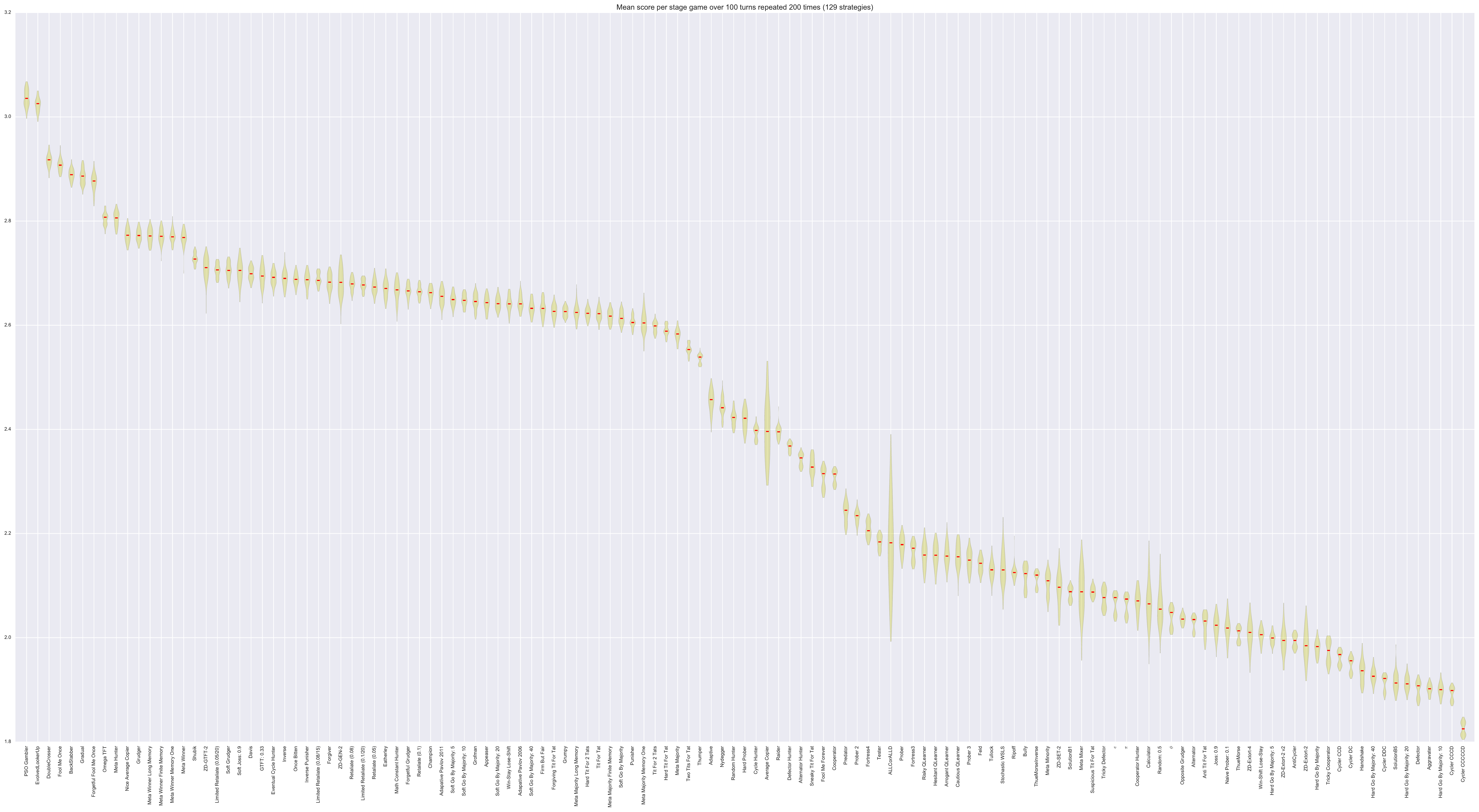}
    \caption{Results from the library tournament (2016-06-13)}
    \label{fig:tournament}
\end{figure}

The current winning strategy is new to the research literature: Looker Up. This
is a strategy that maps a given set of states to actions. The state space is
defined generically by \(m, n\) so as to map states to actions as shown
in~(\ref{equ:example_of_lookerup}).

\begin{equation}
    (\underbrace{(C, D, D, D, C, D, D, C)}_{m\text{ first actions by opponent}},
    \overbrace{((C, C), (C, C))}^{n\text{ last pairs of actions}}) \to D
    \label{equ:example_of_lookerup}
\end{equation}

The example of (\ref{equ:example_of_lookerup}) is an incomplete illustration of
the mapping for \(m=8, n=2\). Intuitively, this state space uses the initial
plays of the opponent to gain some information about its intentions whilst still
taking into account the recent play. The actual winning strategy is an instance
of the framework for \(m=n=2\) for which a particle swarm algorithm has been
used to train it. The second placed strategy was trained with an evolutionary
algorithm \cite{Jones2015, George2016}.
In \cite{Knight2015} experiments are described that evaluate how the second
placed strategy behaves in environments other than those in which it was trained
and it continues to perform strongly.

There are various other insights that have been gained from ongoing open
research on the library, details can be found in \cite{Harper2015}. These include:

\begin{itemize}[noitemsep,topsep=0pt]
    \item A closer look at zero determinant strategies, showing that
		extortionate strategies obtain a large number of wins: the number of
		times they outscore an opponent during a given match. \textit{However} these do
		not perform particularly well from the overall tournament ranking point of
		view. This is relevant given the findings of \cite{Stewart2012} in which zero
		determinant strategies are shown to be able to perform better than any other
		strategy. This finding extends to noisy tournaments (which are also implemented
		in the library).
    \item This negative relationship between wins and performance does not
        generalise. There are some strategies that perform well, both in terms
        of matches won and overall performance: Back stabber, Double crosser,
        Looker Up, and Fool Me Once. These strategies continue to perform well in noisy
        tournaments, however some of these have knowledge of the length of the
        game (Back stabber and Double crosser). This is not necessary to rank
        well in both wins and score as demonstrated by Looker Up and Fool Me
        Once.
    \item Strategies like Looker Up and Meta Hunter seem to be generally
        cooperative yet still exploit naive strategies. The Meta Hunter strategy
        is a particular type of Meta strategy which uses a variety of other
        strategy behaviours to choose a best action. These strategies perform
        very well in general and continue to do so in noisy tournaments.
\end{itemize}

\section*{Conclusion}

This paper has presented a game theoretic software package that aims to address
reproducibility of research into the Iterated Prisoner's Dilemma. The open
nature of the development of the library has lead rapidly to the inclusion of
many well known strategies, many novel strategies, and new and recapitulated
insights.

The capabilities of the library mentioned above are not at all comprehensive, a
list of the current abilities include:

\begin{itemize}[noitemsep,topsep=0pt]
    \item Noisy tournaments.
    \item Tournaments with probabilistic ending of interactions.
    \item Ecological analysis of tournaments.
    \item Moran processes.
    \item Morality metrics based on \cite{Singer-Clark2014}.
    \item Transformation of strategies (in effect giving an infinite number of
        strategies).
    \item Classification of strategies according to multiple dimensions.
    \item Gathering of full interaction history for all interactions.
    \item Parallelization of computations for tournaments with a high
        computational cost.
\end{itemize}

These capabilities are constantly being updated.

\section*{Acknowledgements}

The authors would like to thank all contributors. Also, they thank
Robert Axelrod himself for his well wishes with the library.

\section*{Competing interests}

The authors declare that they have no competing interests.

\printbibliography
%\bibliographystyle{plain}
%\bibliography{references.bib}
\end{document}